\documentclass[journal]{IEEEtran}
\usepackage{cite}
\usepackage{amsmath,amssymb,amsfonts}
\usepackage{algorithm}
\usepackage{algpseudocode}
\usepackage{graphicx}
\usepackage{textcomp}
\usepackage{xcolor,soul}
\setstcolor{red}
\sethlcolor{blue}
\usepackage{bm}
\usepackage{subfigure}
\usepackage{array}
\usepackage{tabularx}
\newcolumntype{Y}{>{\centering\arraybackslash}X}
\newcolumntype{K}[1]{>{\centering\arraybackslash}m{#1}}
\newcolumntype{L}[1]{>{\centering\arraybackslash}p{#1}}
\usepackage{makecell}
\usepackage{array}
\usepackage{multirow}
\usepackage{hyperref}
\usepackage{booktabs}
\usepackage{colortbl}
\usepackage{threeparttable}
\ifCLASSINFOpdf

\begin{document}

\title{Xpikeformer: Hybrid Analog-Digital Hardware Acceleration for Spiking Transformers}
\author{Zihang~Song
        Prabodh Katti,
        Osvaldo~Simeone,
        and~Bipin~Rajendran
\thanks{The authors are with the Centre for Intelligent Information Processing
Systems (CIIPS), Department of Engineering, King’s College London,
WC2R 2LS London, U.K.  (zihang.song@kcl.ac.uk, prabodh.katti@kcl.ac.uk, osvaldo.simeone@kcl.ac.uk, bipin.rajendran@kcl.ac.uk)}
\
\thanks{This work is supported in part by the European Union’s Horizon Europe project CENTRIC (101096379), the Engineering and Physical Science Research Council (EPSRC) project (EP/X011852/1) and by Open Fellowships of the EPSRC (EP/W024101/1 and EP/X011356/1).}
}

\maketitle

\begin{abstract}
The integration of neuromorphic computing and transformers through spiking neural networks (SNNs) offers a promising path to energy-efficient sequence modeling, with the potential to overcome the energy-intensive nature of the artificial neural network (ANN)-based transformers. However, the algorithmic efficiency of SNN-based transformers cannot be fully exploited on GPUs due to architectural incompatibility. This paper introduces Xpikeformer, a hybrid analog-digital hardware architecture designed to accelerate SNN-based transformer models. The architecture integrates analog in-memory computing (AIMC) for feedforward and fully connected layers, and a stochastic spiking attention (SSA) engine for efficient attention mechanisms. We detail the design, implementation, and evaluation of Xpikeformer, demonstrating significant improvements in energy consumption and computational efficiency. Through image classification tasks and wireless communication symbol detection tasks, we show that Xpikeformer can achieve inference accuracy comparable to the GPU implementation of ANN-based transformers. Evaluations reveal that Xpikeformer achieves $13\times$ reduction in energy consumption at approximately the same throughput as the state-of-the-art (SOTA) digital accelerator for ANN-based transformers. Additionally, Xpikeformer achieves up to $1.9\times$ energy reduction compared to the optimal digital ASIC projection of SOTA SNN-based transformers.
\end{abstract}

\IEEEpeerreviewmaketitle

\section{Introduction}

The transformer architecture has revolutionized various fields within artificial intelligence, becoming the backbone of many state-of-the-art (SOTA) models used in natural language processing (NLP) \cite{vaswani2017attention}, computer vision (CV) \cite{dosovitskiy2021image}, and wireless communications {\cite{zecchin2023context,rajagopalan2023transformers,gunduz2023transformer}}, but it also imposes a significant burden on computing and memory resources. This is due to their vast number of parameters and the large amount of matrix-vector and matrix-matrix multiplications required by dense feed-forward layers as well as the attention mechanism. 

\begin{figure}[t]
    \centering
    \subfigure[]{\includegraphics[width=1\linewidth]{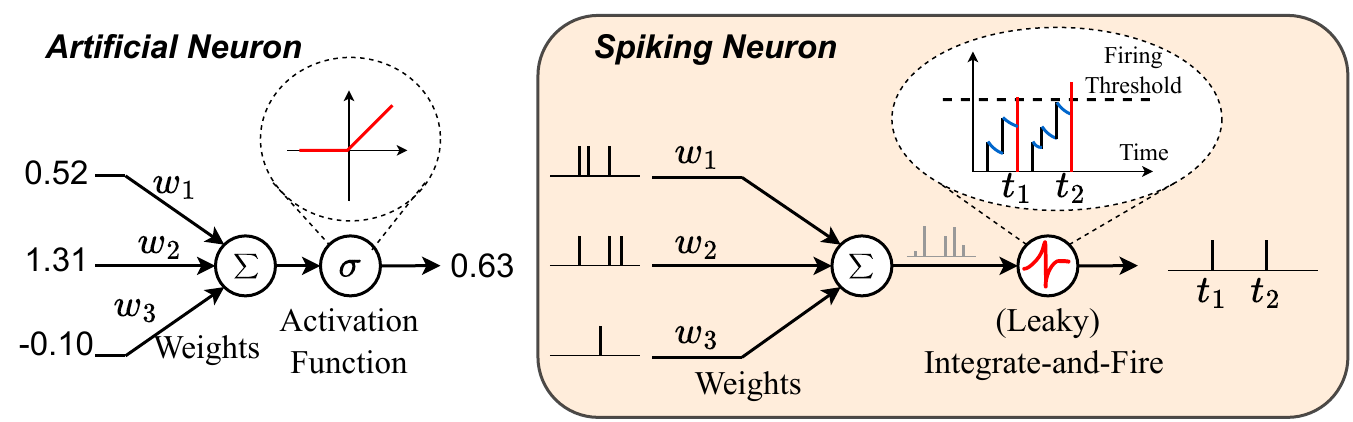}\label{fig_neurons}}
    \subfigure[]{\includegraphics[width=1\linewidth]{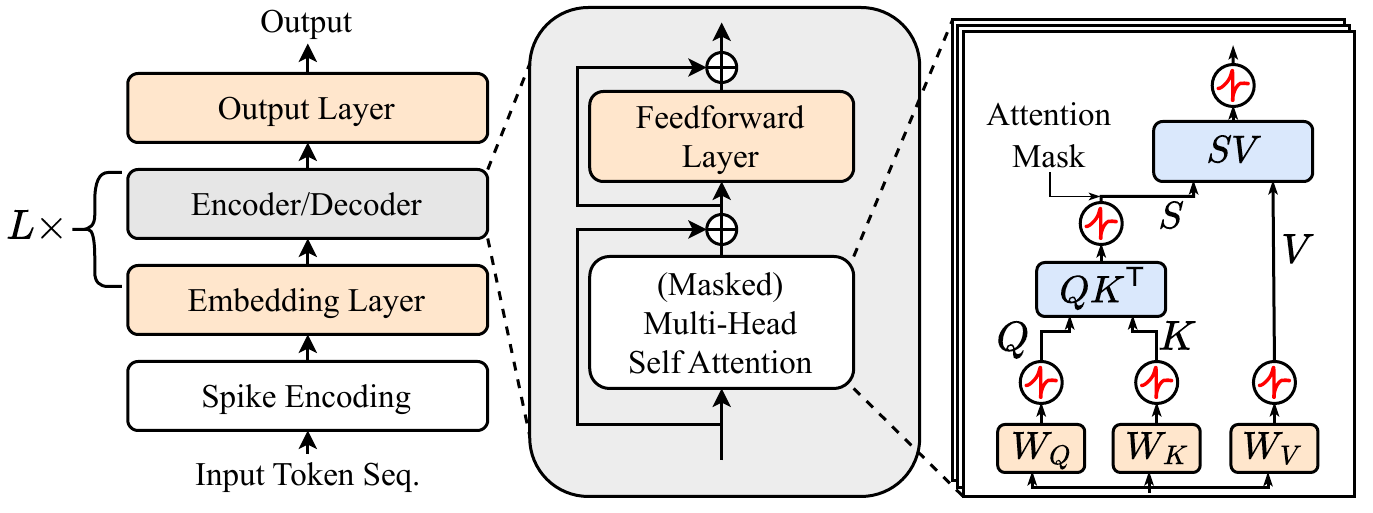}\label{fig_stranformer}}
    \caption{(a) Comparison of an artificial neuron (left) and a spiking neuron (right). (b)  A simplified block diagram of a spiking transformer.}
    \label{fig:ann-snn}
    \vspace{-5mm}
\end{figure}

The integration of \textit{neuromorphic computing} with transformer architectures, enabled by spiking neural networks (SNNs), represents a promising frontier for achieving energy-efficient and biologically-plastic sequence modeling \cite{she2022spikeformer,9319553,10318216,9452789}. As illustrated in Fig.~\ref{fig_neurons}, neuromorphic computing uses sparse, temporally-encoded spikes for neuronal signaling \cite{maass1997networks}. This spike-based signaling allows spiking neurons to aggregate activations using efficient accumulate (AC) operations, in contrast to artificial neurons in traditional artificial neural networks (ANNs), which process continuous-valued activations through multiply-and-accumulate (MAC) operations. \emph{SNN-based transformers}, or \textit{spiking transformers}, as shown in Fig.~\ref{fig_stranformer}, adopts spiking neurons in feed-forward operations while incorporating spiking-based self-attention mechanisms, achieving higher computing efficiency in sequence modeling tasks compared to ANN-based transformers \cite{yao2023spike,xu2024spikezip,zhou2022spikformer,zhou2023spikingformer,zhou2024spikformer,bal2023spikingbert,zhu2023spikegpt,she2022spikeformer,zhang2022spiking,yao2024spike}.

Although spiking transformers are algorithmically efficient due to their simplified computations, they face significant inefficiencies on general-purpose computing platforms like GPUs. These inefficiencies stem from a fundamental mismatch between the event-driven, sparse nature of spiking algorithms and GPU architectures, which are optimized for dense, high-precision workloads. First, spiking transformers operate on binary data, while GPUs are designed for high-precision arithmetic (e.g., FP16 or FP32), leading to poor resource utilization and suboptimal energy efficiency. Second, the temporal dimension of spiking transformers introduces substantial data communication overhead, as frequent memory access for intermediate results significantly increases latency and energy consumption. While optimization efforts for spiking transformers have primarily focused on algorithmic and software improvements, the lack of dedicated hardware accelerators remains a critical gap in the field \cite{song2024stochastic}.

In recent years, analog in-memory computing (AIMC) using non-volatile memory (NVM) devices has emerged as a promising mixed-signal approach for accelerating both ANNs and SNNs \cite{sebastian2020memory}. AIMC enables matrix-vector multiplications (MVMs) to be performed in constant time, $O(1)$, directly within \textit{crossbar arrays} of NVM devices. This is achieved by storing the matrix elements as conductances of the NVM devices and by applying input vector values as electrical voltages to the crossbar input lines and reading the resulting electrical currents in the crossbar output lines, leveraging Kirchhoff's laws. Several AIMC-based ANN transformer accelerators have been proposed, significantly improving the efficiency of ANN transformers \cite{yang2020retransformer,yang2022fullcircuit,sridharan2023x}. However, due to architectural differences between ANN-based transformers and spiking transformers, these accelerators cannot be directly applied to the latter. 

Additionally, AIMC is primarily effective for accelerating inference in feedforward and fully connected layers, where weights are programmed into the crossbar once and reused during forward propagation. However, AIMC is less suited for attention mechanisms, which involve matrix multiplications between two dynamic matrices, requiring frequent data writes to NVM crossbars. This introduces latency, increases energy overhead, and is limited by NVM endurance \cite{merced2016repeatable, yang2020retransformer}. As a result, digital acceleration is better suited for attention. Although these matrix multiplications can be simplified to masked additions in spiking domain owing to the binary nature of the multiplicands \cite{zhou2022spikformer}, executing them on GPUs—designed for high-precision arithmetic—remains resource-inefficient.

In this paper, we introduce \textit{Xpikeformer}, the first hardware accelerator specifically designed for spiking transformers. The architecture integrates two specialized computing frameworks to optimize performance and energy efficiency. For static-weight operations, including feed-forward, embedding, and fully connected layers, Xpikeformer employs AIMC for efficient in-memory computation and weight storage. For the attention mechanism, it utilizes stochastic computing with lightweight logical operations and a streaming dataflow, allowing rapid processing without intermediate result storage. Additionally, we introduce a simulation framework to train Xpikeformer and evaluate its accuracy performance. The key contributions of this work are as follows:

\begin{itemize}
    \item We design an AIMC engine to efficiently handle feed-forward, embedding, and fully connected layers in spiking transformers without storing non-binary pre-activations, significantly reducing data transfer overhead.
    \item We propose a software-hardware co-design methodology for a spike-based attention architecture, termed stochastic spiking attention (SSA), which leverages stochastic computing to execute matrix multiplications in the attention mechanism of spiking transformers using lightweight logic gates instead of heavy arithmetic units.  
     \item Xpikeformer achieves a $13\times$ energy reduction at approximately the same throughput as the SOTA digital accelerator for ANN-based transformer. As the first hardware accelerator design for spiking transformers, Xpikformer also showcases $1.8$--$1.9\times$ energy reduction compared to the ideal digital ASIC projection of the SOTA spiking transformer software design. Compared to GPU implementations, Xpikeformer achieves a $2.18\times$ speedup over traditional ANN-based transformers and a $6.85\times$ speedup over spiking transformers.
\end{itemize}

This paper is structured as follows: Section \ref{se_pre} provides an overview of the preliminaries. Section \ref{se_relatedwork} reviews related work on hardware accelerators for ANN-based and spiking transformers. Section \ref{se_archi} details the design and implementation of the Xpikeformer architecture. Section \ref{se_train} discusses the training and inference methodologies. Section \ref{se_acc} presents the accuracy evaluation of the proposed system, and Section \ref{se_energy} assesses its efficiency. Finally, Section \ref{se_conclusion} concludes the paper.

\section{Preliminaries}\label{se_pre}
\begin{table*}[t]
    \centering
    \caption{Comparison of operations in ANN transformers, a typical SOTA spiking transformer architecture, and the spiking transformer introduced in this work (GeLU = Gaussian Error Linear Unit, LIF = Leaky Integrate-and-Fire, BNL = Bernoulli Neuron Layer)}
    \begin{tabular}{cccc}
    \hline
        \textbf{Operations} & \textbf{ANN}\cite{vaswani2017attention} & \textbf{SNN}\cite{zhou2022spikformer,yao2023spike,zhou2024spikformer} & \textbf{SNN (Xpikeformer)} \\ \hline 
       \rule{0pt}{3ex}   \(Q\), \(K\), \(V\) Generation & Linear & Linear + LIF & Linear + LIF\\ 
       \rule{0pt}{3ex}Attention & $\text{softmax}\left(\cfrac{QK^T}{\sqrt{d_k}}\right)V$ & $\text{LIF}(\text{LIF}({Q^t{K^t}^{\mathsf{T}}})V^t)$ & $\text{BNL}(\text{BNL}(Q^t {K^t}^{\mathsf{T}}) V^t )$\\
       \rule{0pt}{3ex} Feedforward & $W_2\left(\text{GeLu}(W_1X)\right)$ & $\text{LIF}\left(W_2\text{LIF}({W_1X^t})\right)$ & $\text{LIF}\left(W_2\text{LIF}({W_1X^t})\right)$\\
        \rule{0pt}{3ex}\makecell[tc]{Inter-layer\\Normalization} & LayerNorm($X$) & None & None\\ 
        \hline
    \end{tabular}
    \vspace{-2mm}
    \label{tab:ann_snn_transformers}
\end{table*} 
\subsection{Spike Encoding and Spiking Neurons}
{One essential difference between SNNs and ANNs is the {information encoding method} \cite{yang2023nadol}. In ANNs, activations are represented by real values \(x\). In contrast, SNNs encode activations in binary temporal sequences \(s[t] \in \{0,1\}\) for \(t = 1, 2, \ldots, T\), known as \emph{spike trains}, where \(T\) denotes the spike encoding length.} A commonly utilized spike encoding scheme to map a real value \(x \in [0, 1]\) {into} a spike train is \emph{Bernoulli rate coding}, whereby each bit in the encoded sequence independently takes the value `1' with probability \(x\) and `0' with probability \(1 - x\), {which is expressed as}:
\begin{equation}\label{eq_bernoulli}
    s[t] \sim \text{Bern}(x) \text{ for } t = 1, 2, \ldots, T.
\end{equation}

Fig.~\ref{fig_neurons} illustrates the comparative differences between a spiking neuron and an artificial neuron. Traditional artificial neurons in ANNs aggregate weighted real-valued inputs and apply static activation functions (e.g., sigmoid, tanh, or ReLU) to produce the activation. In contrast, a spiking neuron integrates the weighted sum of input spikes using the leaky integrate-and-fire (LIF) membrane model over discrete time steps. The membrane potential \( V^t \) of an LIF neuron at time step \( t \) evolves as
\begin{equation}\label{eq:V}
\begin{aligned}
V^t &= \beta V^{t-1} + I^t,\\
\end{aligned}
\end{equation}
where \( \beta\in(0,1] \) represents the leak factor determining the rate of decay of the membrane potential over time, and \( I^t \) is the input at time step \( t \). When \( V^t \) reaches or exceeds a threshold voltage \( V_{\text{thresh}} \), the neuron fires a spike, i.e., $O^t=1$, and resets the potential $V^t$ to $0$. Otherwise, the neuron remains silent, producing $O^t=0$, i.e., 
\begin{equation}\label{eq:th}
\text{if } V^t \geq V_{\text{thresh}} \text{ then }
\begin{cases}
O^t = 1,\\
V^t = 0,
\end{cases}
\text{ else } O^t = 0 \text{ and \eqref{eq:V}}.
\end{equation}

\subsection{Stochastic Computing}
An important advantage of Bernoulli rate coding is that it supports \emph{stochastic computing}, a computational paradigm that employs Bernoulli bit streams \cite{Alaghi2013Survey}. Consider two Bernoulli spike trains \( s_{1}[t] \) and \( s_{2}[t] \) with spiking probabilities \( p\left(s_1[t]=1\right)=x_1 \) and \( p\left(s_2[t]=1\right)=x_2 \). The multiplication of real-valued \( x_1 \) and \( x_2 \) can be achieved by applying the bit-wise logic AND (\(\land\)) operation on their stochastic representations \( s_{1}[t] \) and \( s_{2}[t] \), which is denoted as:
\begin{equation}
   s_{\text{out}}[t] = s_{1}[t] \;\land\; s_{2}[t],
\end{equation}
since \( p\left(s_{\text{out}}[t]=1\right)= p\left(s_1[t]=1\right) \cdot p\left(s_2[t]=1\right) = x_1 x_2 \). In this paper, we utilize stochastic computing to help accelerate the matrix multiplications within spiking transformers, as will be detailed in Section \ref{se1_ssa}.

\subsection{Transformers and Spiking Transformers}
\subsubsection{ANN-based Transformer}
The typical architecture of an ANN-based transformer model is based on stacks of encoder layers \cite{devlin2018bert,dosovitskiy2021image}, decoder layers \cite{brown2020language}, or both \cite{vaswani2017attention}. Each encoder or decoder layer is composed of a multi-head self-attention (MHSA) mechanism and a feedforward network with residual connections and layer normalization. In models that include both encoder and decoder stacks, the decoder layers also contain an additional multi-head self-attention mechanism that attends to the encoder's output.

The {MHSA} mechanism in transformers allows the model to compute attention scores that determine the relevance of each token in the input sequence with respect to every other token. This enables the model to weigh and integrate information from different positions in the sequence, capturing various contextual dependencies. Multiple heads attend to the sequence elements on different subspaces simultaneously. For each head, the input token embeddings are projected using projection matrices $W_Q$, $W_K$, and $W_V$ to get the query $Q$, key $K$, and values $V$, respectively. The dot-product attention is then calculated as follows:
\[
\text{attention}(Q, K, V) = \text{softmax}\left(\frac{QK^T}{\sqrt{d_k}}\right) V,
\]
where $d_k$ is the dimension of key vectors within each head.

\subsubsection{Spiking Transformers} \label{se1_spikingtransformers}
Spiking transformers are a novel adaptation of traditional ANN-based transformers that integrates principles from SNNs {\cite{yao2023spike,xu2024spikezip,zhou2022spikformer,zhou2023spikingformer,zhou2024spikformer,bal2023spikingbert,zhu2023spikegpt,she2022spikeformer,zhang2022spiking,yao2024spike}}. The typical architecture of a spiking transformer is shown in Fig. \ref{fig_stranformer}. In this architecture, input tokens are first encoded into temporal spike trains through a spike encoding layer. The subsequent architecture resembles that of ANN transformers, with the operations of ANN-based transformers and spiking transformers compared in Table \ref{tab:ann_snn_transformers}. The operation of the proposed Xpikeformer is also listed in Table \ref{tab:ann_snn_transformers} and will be detailed in Section \ref{se1_ssa}.

In a spiking transformer, the feedforward, embedding, and fully connected layers are implemented using spiking neurons in spiking transformers. A key distinction in MHSA lies in how the token representations \(Q\), \(K\), and \(V\) are generated and processed. In ANN-based transformers, the representations \(Q\), \(K\), and \(V\) are obtained through linear projections applied to the token embeddings $X$. In contrast, spiking transformers apply LIF neurons after linear projections, resulting in spike-encoded representations \(Q^t\), \(K^t\), and \(V^t\). Compared to the attention score computation \( S = \text{softmax}(QK^{\mathsf{T}}/\sqrt{d_K})\) in ANN-based transformers, spiking transformers first perform matrix multiplication between matrices $Q^t$ and $K^t$ independently for each time step $t$, then apply a LIF neuron to the product, expressed as $S^t= \text{LIF}(Q^t{K^t}^{\mathsf{T}})$. Another notable difference in spiking transformer MHSA is the absence of softmax. ANN-based transformers use the softmax function to normalize the attention scores, ensuring they sum to one. However, in spiking transformers, softmax is impractical due to the binary nature of attention scores and the need for temporal and sparse processing. Instead, spiking neurons regulate the activation through their firing dynamics \cite{zhou2022spikformer}. Finally, the attention score $S^t$ is multiplied with  \( V^t \) at each time step $t$, followed by a LIF neuron to produce the final attention output.

\subsubsection{Efficiency Challenges}
Spiking transformers achieve accuracy comparable to their ANN counterparts in NLP \cite{bal2023spikingbert}, CV \cite{zhou2024spikformer}, and wireless communication \cite{song2024neuromorphic}. Estimated on neuromorphic hardware assumptions, their energy efficiency suggests up to an order-of-magnitude reduction in computational energy by replacing matrix multiplications with addition and integer multiplication. However, deploying spike-based models on general-purpose platforms like CPUs and GPUs results in significant energy inefficiencies due to:

\begin{itemize}
    \item \textit{Data Communication Overhead}: Frequent memory access for intermediate results in feedforward and MHSA layers introduces overhead, especially when the spike encoding length \(T\) exceeds GPU bit widths. Non-binary pre-activations generated at each time step further amplify this overhead, increasing transmission requirements by a factor of \(T\) compared to equally quantized ANNs.
    
    \item \textit{Precision Mismatch}: SNNs operate on binary data, whereas CPUs and GPUs are optimized for high-precision computation (e.g., FP32/64), leading to inefficiencies. While some GPUs support INT8, even this precision exceeds the needs of binary spiking signals. FPGA-based spiking attention accelerators may address this issue, but remain underexplored \cite{isik2023hpcneuronet}.
\end{itemize}

\subsection{Analog In-Memory Computing (AIMC)}\label{se1_crossbar}
AIMC enables fast and energy-efficient MVMs using synaptic arrays (SAs). An SA primarily comprises an NVM crossbar. The crossbars consist of NVM devices, which are NVM elements positioned at row-column intersections, where each element can be programmed to different conductance levels, representing matrix values. This structure allows the crossbar to store matrices with significantly greater area efficiency than traditional digital buffers. To perform an MVM, the input vector is encoded as voltages applied to the crossbar rows. Ohm’s Law governs the current flow through each NVM device, while Kirchhoff’s Current Law ensures the correct summation of currents at the columns, yielding the dot-product result. Beyond the crossbar, an SA includes readout units, typically comprising successive approximation-register (SAR) analog-to-digital converters (ADCs) coupled with multilevel current-mode sense amplifiers. These are integrated with accumulation circuits (e.g., adders and registers) and various peripheral components, including decoders, multiplexers, switch matrices, and buffers.

\begin{figure}[t]
    \centering
    \includegraphics[width=0.9\linewidth]{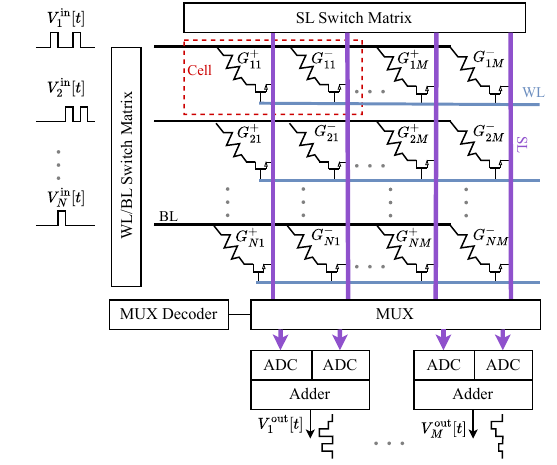}
    \caption{Illustration of a typical implementation of synaptic array \cite{peng2020dnn+} with spike-encoded signals as input.}
    \label{fig:crossbar}
\end{figure}

A typical implementation of an SA, introduced by \cite{peng2020dnn+}, is illustrated in Fig.~\ref{fig:crossbar}. In this setup, every two devices in the crossbar form a differential pair cell, enabling the representation of positive and negative matrix values. The input spike-encoded signals are applied through the bit lines (BLs), while the results are read from the source lines (SLs). The word lines (WLs) and SLs control the switching of NVM devices via transistors. During the weighted sum operation, all cells are activated when all WLs are turned on. Input vector voltages are applied to the BLs, and the resulting weighted sum currents are read out in parallel through the SLs. These currents are then digitized by multiple readout units concurrently. To optimize the balance between area, power, and latency, the readout units are shared among a predefined number of crossbar columns and time-multiplexed under the control of a multiplexer (MUX).

In a spiking transformer, input neuron activations are represented by spike trains, which consist of multiple cycles of input voltage signals to the BLs. Since these input signals encode 1-bit binary data at each time step, analog voltage representation is unnecessary, eliminating the need for digital-to-analog converters (DACs) during inference. However, DACs are still required in hardware for the initial programming of NVM conductances to the appropriate analog levels corresponding to the matrix elements.

\section{Related Work}\label{se_relatedwork}
\subsection{ANN Transformer Accelerators}
\subsubsection{Digital Accelerators}
The development of digital (fully CMOS-based) accelerators for ANN-based transformers primarily focuses on enhancing computational efficiency. Various strategies have been explored to achieve this, including mixed-precision techniques \cite{shao2023efficient}, algorithm-specific quantization \cite{marchisio2023swifttron}, and adaptive quantization methods based on input data characteristics \cite{ham2020accelerating}. One approach decomposes major operations into a single dot product primitive, enabling unified and efficient execution, significantly improving throughput \cite{wang2022row}. Another optimization involves a dedicated data path for attention mechanisms, reducing latency and further enhancing throughput \cite{ham2020accelerating}. Additionally, an approach to decoding inefficiencies skips redundant computations and adopts a sparse matrix format, maximizing MAC utilization \cite{park2020optimus}. Despite these advancements, existing solutions remain inadequate in addressing data transfer overhead, which continues to be a major bottleneck.

\subsubsection{AMIC-Integrated Accelerators}
Advancements in AIMC have enhanced efficiency and performance in ANN transformer accelerators. A hybrid approach employs AIMC for feedforward layers while executing MHSA on general-purpose units, achieving software-equivalent accuracy for NLP tasks but retaining energy inefficiencies due to high-precision digital computations \cite{okazaki2022analog,spoon2021toward}. 

Efforts to implement MHSA with AIMC require online programming to store intermediate results (\(K\) and \(V\)) into crossbars, introducing significant latency, power consumption, and impracticality due to NVM wear \cite{yang2020retransformer}. Circuit-based implementations replace online programming with analog modules for dot-product attention, softmax, and activation functions, aligning with crossbar feedforward layers. However, analog circuits suffer from non-linearity, noise, and higher power demands compared to digital designs \cite{yang2022fullcircuit}.

Alternative architectures explore near-memory computing (NMC) or digital in-memory computing (DIMC) using SRAM for MHSA workloads. An 8T-SRAM-based hybrid accelerator reduces leakage and employs sequence-blocking dataflows to enhance hardware utilization and reduce execution time \cite{zhou2022transpim,li2023h3datten,sridharan2023x}. However, these designs still require writing key and value matrices to SRAM during inference, adding latency and energy overhead, while also relying on compute-intensive softmax operations.

\subsection{Advances Towards Efficient Spiking Transformers}
Despite extensive research on hardware accelerators for SNNs, particularly for LIF models \cite{bouvier2019spiking,carpegna2022spiker,liu2022fpga}, efforts to accelerate spiking transformers have remained largely algorithmic and software-focused. One pioneering study in this domain proposes a spiking MHSA that captures sparse visual features using spike-form \(Q\), \(K\), and \(V\) without employing softmax, achieving ANN-comparable accuracy in image classification tasks \cite{zhou2022spikformer,zhou2024spikformer}. Another significant contribution presents a hardware-friendly spike-driven residual learning architecture that eliminates non-spike computations from residual connections \cite{zhou2023spikingformer}. Building on these advancements, \cite{yao2024spike} further reduces computational energy by replacing multiplications with sparse addition operations in MHSA. However, these studies remain at the software and algorithmic level, making them inefficient on general-purpose computing platforms like CPUs and GPUs \cite{isik2023hpcneuronet}. Furthermore, none have addressed the data transfer overhead, leaving a critical aspect of efficiency improvement unexplored.

\section{Xpikeformer Architecture}\label{se_archi}

\begin{figure}[t]
    \centering
    \includegraphics[width=0.8\linewidth]{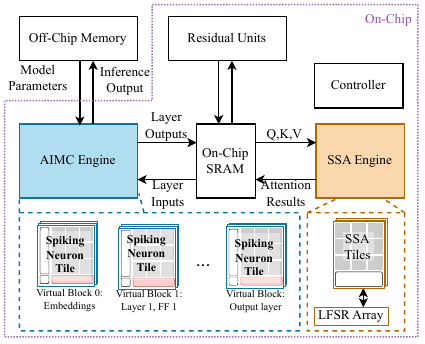}
    \caption{The overall system architecture of Xpikefomer.}
    \label{fig:system}
\end{figure}

Xpikeformer is a hybrid analog-digital architecture designed for spiking transformers. As shown in Fig.~\ref{fig:system}, Xpikeformer integrates two specialized computing engines for optimized performance. The \textit{AIMC engine} handles feed-forward, embedding, and fully connected layers using spiking neuron tiles, with NVM crossbars for efficient weight storage. The \textit{SSA engine} implements the MHSA mechanism in the spiking domain leveraging stochastic computing. This CMOS-based design enhances efficiency by replacing arithmetic units with logical AND gates.

\subsection{AIMC Engine}\label{se1_aimcengine}
The AIMC engine is composed of multiple spiking neuron tiles, which are virtually grouped to ensure sufficient memory for storing weights in each feedforward or fully connected layer. Each tile comprises an input and output buffer, multiple synaptic arrays (SAs), and a group of LIF units. A shared DAC unit is integrated into each spiking neuron tile to facilitate the initial weight programming. This DAC unit is bypassed during inference, eliminating unnecessary overhead.

\subsubsection{NVM Crossbars}
Each SA includes a fixed-sized NVM crossbar array along with multiplexed readout units that allow MVMs to be performed in constant time, $O(1)$, directly within the memory, as introduced in Section \ref{se1_crossbar}. This eliminates the need for energy-intensive CMOS arithmetic operations in conventional digital implementations, significantly reducing computational energy consumption. We impose a restriction on the size of crossbar arrays, because excessively large sizes result in increased interconnect resistance and capacitance, higher power consumption, signal degradation, and increased complexity of control circuitry. Consequently, the weights of a feedforward or fully connected layer are most likely distributed across multiple SAs or even several spiking neuron tiles.

We choose the phase change memory (PCM) device to implement the crossbars in SAs due to its multi-level cell capability (up to 4 bits per device), high endurance, fast read/write speed, and compatibility with CMOS technology \cite{wu201840nm,wong2010phase,mehonic2020memristors}. Table \ref{tab:parameters} summarizes the detailed SA configuration parameters used in this work. Nevertheless, as an analog device, PCM is susceptible to thermal noise, $1/f$ noise, device variability, and quantization effects. A key challenge is conductance drift, where the stored conductance of a memory cell changes over time after programming. This drift can degrade neural network performance by affecting the precision and reliability of stored weights. Methods to address the effects caused by these nonidealities will be introduced in Section \ref{se_train}.

\begin{table}
    \centering
    \caption{Xpikeformer Configuration of Synaptic Arrays}
    \begin{tabular}{cc}
         \toprule 
        \textbf{Parameter} & \textbf{Value}\\  \toprule 
        Resistive device & PCM\\
        Conductance Resolution & 4 bits \\
        Weight Resolution & 5 bits \\
        \# devices per cell & 2 \\
        Crossbar dimension (by cell)& $128\times 128$ \\
        ADC resolution & 5 bits \\
        ADC sharing ratio & 8 \\
         \toprule 
     \end{tabular}    
    \label{tab:parameters}
\end{table}

\subsubsection{Weight Mapping and Tile Architecture}
\begin{figure}[t]
    \centering
    \includegraphics[width=0.9\linewidth]{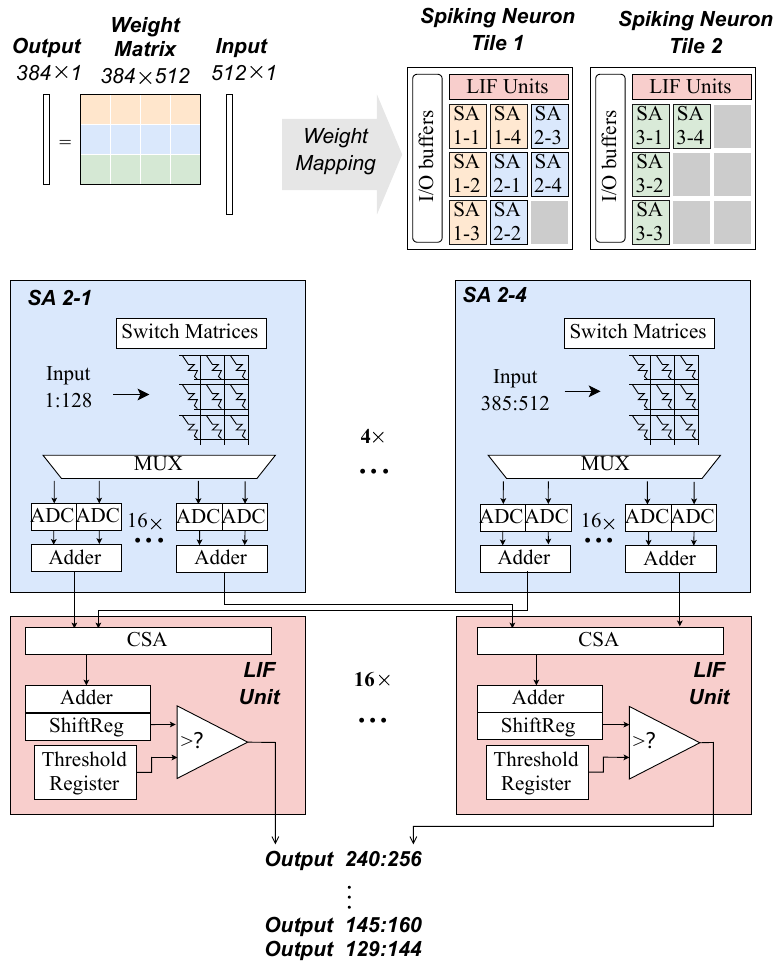}
    \caption{A illustration of the row-block-wise mapping strategy.}
    \label{fig:mapping}
\end{figure}

To efficiently manage this distribution and avoid the overhead of storing non-binary pre-activations, we adopt a \textit{row-block-wise mapping strategy}. For example, with a crossbar size of 128$\times$128 cells and a weight matrix of 384$\times$512, the weight matrix must be divided into twelve 128$\times$128 submatrices. As shown in Fig.~\ref{fig:mapping}, the row-block-wise mapping strategy ensures that the same row of submatrices is mapped to the SAs within a single spiking neuron tile. For instance, rows 129-256 of the weight matrix are mapped to four SAs (2-1 to 2-4) within one spiking neuron tile. The input 512$\times$1 vector is then split into four 128$\times$1 subvectors and fed into SA 2-1 to SA 2-4, respectively. 

At each SA column output of an SA, only a local sum of the vector dot product is obtained. To avoid storing this sum, the outputs from the same column of SA 2-1 to SA 2-4 are routed to a shared LIF unit, where they are accumulated using a carry-save adder (CSA) to produce a single pre-activation value. This pre-activation is then added to the current membrane potential via an adder. 

The new membrane potential overwrites the previous one in a shift register. A comparator then compares this value against a threshold stored in the threshold register. If the membrane potential exceeds the threshold, the comparator generates a `1' and resets the shift register. At each time step, the shift register undergoes a right shift by one bit, representing a leak factor of $\beta = 0.5$. The comparator output, which acts as the spiking neuron’s output, is sent to the output buffer and subsequently stored in shared SRAM.

The row-block-wise mapping strategy is compatible with ADC sharing in SAs. As shown Fig.~\ref{fig:mapping}, the sharing ratio is 8, meaning each SA has 128$/$8$=$16 readout units. Consequently, SA 2-1 to SA 2-4 are connected to 16 LIF units. The address decoding strategy of the MUX decoder is identical across all SAs to ensure alignment between local sums. During each MUX decoding cycle, 16 out of 128 output features are obtained.

This row-block-wise mapping strategy minimizes data movement latency by directly routing local sums from multiple crossbars to LIF units, where they are accumulated and processed without intermediate buffering. This design reduces memory access bottlenecks, ensuring low latency and high energy efficiency.

\subsection{SSA Engine}\label{se1_ssa}

Stochastic spiking attention (SSA) is a software-hardware co-designed spike-based attention mechanism optimized to minimize computing and memory access overhead. Unlike ANN-based attention, which relies on computationally intensive real-valued matrix multiplications, and SNN-based attention, which often uses masked integer addition, SSA utilizes stochastic computing to enable ultra-lightweight binary operations. This is made possible by the binary and stochastic nature of the queries \(Q\), keys \(K\), and values \(V\), which are encoded using Bernoulli neurons. We introduce the SSA engine following a bottom-up approach.

\subsubsection{SSA Algorithm}
As introduced in Table \ref{tab:ann_snn_transformers}, the spiking attention mechanism introduced in \cite{zhou2022spikformer} evaluate the attention block as:
\begin{equation}
    \text{Attention}(Q^t, K^t, V^t) = \text{LIF}(\text{LIF}(Q^t {K^t}^{\mathsf{T}}) V^t ),
\end{equation}
where LIF($\cdot$) represents the per-entry application of LIF neuron as described in Section \ref{se1_spikingtransformers}. The proposed SSA replaces the stateful LIF neurons with the Bernoulli neuron layer (BNL), computing the attention as:
\begin{equation}\label{eq_ssa}
  \text{SSA}(Q^t, K^t, V^t) = \text{BNL}(\text{BNL}(Q^t {K^t}^{\mathsf{T}}) V^t ).
\end{equation}

A Bernoulli neuron is a stateless neuron that takes real-valued input, normalizes the value to a probability in \([0,1]\), and then generates a Bernoulli sequence \eqref{eq_bernoulli} as its output. Using this neuron model, the rate of the output \eqref{eq_ssa} provides an increasingly accurate approximation of real matrix multiplications when \(Q^t\), \(K^t\), and \(V^t\) are Bernoulli-encoded and the encoded sequence is sufficiently long. 

A detailed description of the SSA algorithm is shown in Algorithm \ref{al1}. Owing to the binary nature of the representations $Q^t$, $K^t$, and $V^t$, the matrix multiplications in \eqref{eq_ssa} are realized by lightweight logic AND operations and additions instead of full-precision multiplication units. In step 5, the \(N \times N\) binary attention scores \(S^{t}\) are obtained by first calculating the dot-product between \(Q^t\) and \(K^t\). This is done by accumulating the results of logical AND operations across the dimension \(d_K\). The result is then normalized and used for Bernoulli encoding. Step 9 applies a similar logical AND to compute the matrix multiplication between \(S^t\) and \(V^t\) and generate probabilistically encoded attention result $A^t$.

\begin{algorithm}
\caption{Stochastic Spiking Attention (SSA)}
\begin{algorithmic}[1]
\State \textbf{Input:} $d_K \times N$ binary matrix sequences ${Q}^t, {K}^t, {V}^t$ for $t=1,\ldots,T$, dimension $d_K$ (corresponding to one attention head)
\State \textbf{Output:} $d_K\times N$ binary matrix sequence ${A}^t$ for $t=1,\ldots,T$ (corresponding to one attention head)
\For{$t = 1$ to $T$}
    \For{each $(n,n')$ in $N \times N$}
        \State $S^t_{n,n'} \sim \mathrm{Bern}\left(\frac{1}{d_K} \sum_{d=1}^{d_K} Q^t_{d,n} \land K^t_{d,n'}\right)$ 
    \EndFor
        \State Apply casual mask if using a decoder
    \For{each $(d,n)$ in $d_K\times N $}
        \State\quad $A^t_{d,n} \sim \mathrm{Bern}\left(\frac{1}{N} \sum_{n'=1}^{N} S^t_{n,n'} \land V^t_{d,n'}\right)$
    \EndFor
\EndFor
\end{algorithmic}
\label{al1}
\end{algorithm}

\subsubsection{SSA Tile}
\begin{figure}
    \centering
    \includegraphics[width=\linewidth]{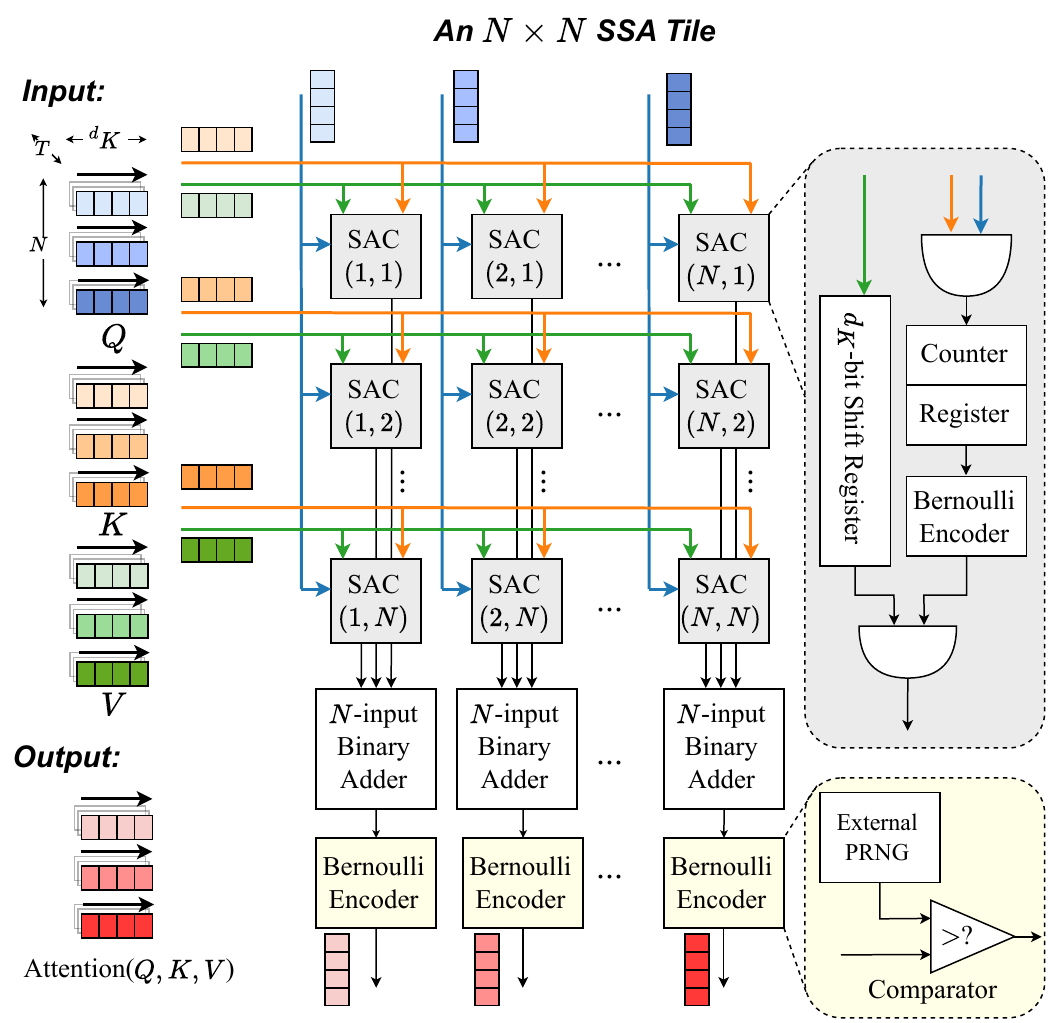}
    \caption{Block diagram of an $N\times N$ SSA tile.}
    \label{fig:ssa}
\end{figure}

We designed a customized hardware architecture, referred to as SSA tiles, optimized for SSA operations. As illustrated in Fig. \ref{fig:ssa}, SSA tiles streamline the computation of spike-based attention by utilizing bitwise logic devices. This design effectively eliminates the data transfer overhead caused by writing and reading intermediate results, significantly enhancing both the energy efficiency and performance of the computations.

{
Each SSA tile is composed of \(N^2\) physical stochastic attention cells (SACs) arranged in an \(N \times N\) array. The \(N^2\) parallelization level in SACs provides several significant benefits. First, it enables the simultaneous pairing of each query with every key-value pair. This is achieved by streaming \(K^t\) and \(V^t\) across columns and \(Q^t\) across rows using 1-bit data buses. Additionally, this architecture allows for the parallel computation of all \(N \times N\) elements of the attention score \(S^t\) matrix. The \((i,j)\)-th element of \(S^t\) is computed within the corresponding \((i,j)\)-th SAC. Given that SOTA transformers for edge AI applications typically operate with \(N\) values ranging from 16 to 128, our design incorporates \(N^2\) physical SAC units to ensure scalability and efficiency.}

To calculate the $(i,j)$-th element of $S^t$, totally \(d_K\) AND operations are needed between the elements of the $i$-th row of $Q^t$ and the $j$-th row of $K^t$. These \(d_K\) AND operations are executed serially in time within the $(i,j)$-th SAC via an AND gate. The summation is realized by counting the AND output using a counter with unsigned INT8 (UINT8) output, accommodating a key dimension \(d_K\) up to \(2^8=256\). After every \(d_K\) clock cycles, the summation is buffered to a Bernoulli encoder, where it is normalized by $d_K$ and used as a probability to generate the Bernoulli sample $S^t_{i,j}$. 

To calculate the attention output $A^t$, the generated attention scores $S^t_{i,j}$ are held in the SAC for $d_K$ clock cycles. The local attention result between $S^t_{i,j}$ and the $j$-th row of $V^t$ is calculated using another AND gate, serving as the output of the SAC. To avoid using external delays for $V^t$, a \(d_K\)-bit shift register operating on a first-in-first-out basis is deployed in each SAC to temporarily buffer \(V^t\) and align it with $S^t$. This allows for the simultaneous streaming of $Q^t$, $K^t$, and $V^t$ within the SSA block, facilitating pipelining over time steps. The summation of the local attention results is achieved by adding the outputs of each column of SACs using an $N$-input binary adder. The sum is then sent to another Bernoulli encoder, where it is normalized by $N$ and used as a probability to generate the Bernoulli samples.  The $i$-th row of $A^t$ is generated sequentially by holding $S^t_{i,j}$ in the $(i,j)$-th SAU when streaming the $j$-th row of $V^t$. As a result of this procedure, the \(i\)th column within the SSA tile sequentially outputs the $i$-th row of the attention result $A^t$. With \(N\) rows of SACs operating in parallel, the entire matrix \(A^t\) is calculated column by column by one SSA tile. 

\begin{figure*}[t]
    \centering
    \includegraphics[width=0.8\linewidth]{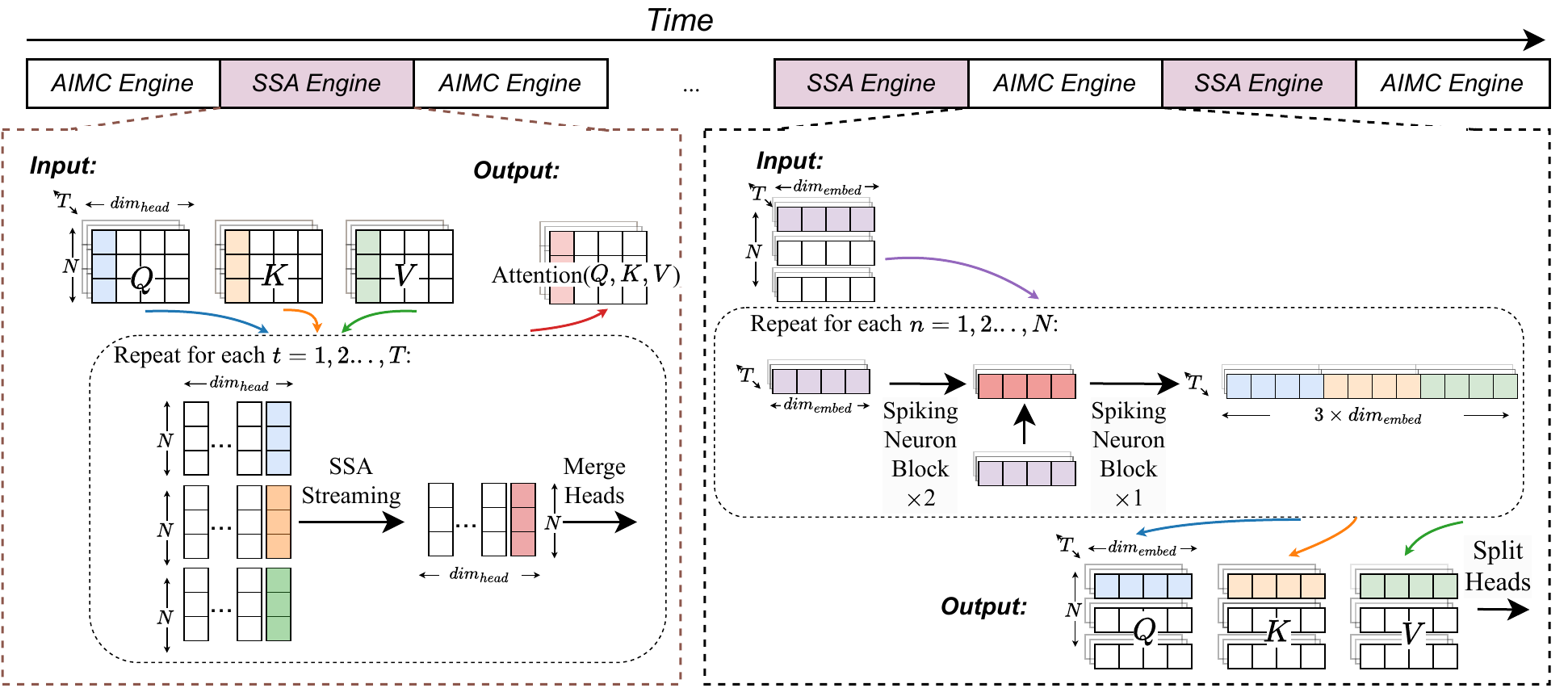}
    \caption{Illustration of Xpikeformer inference dataflow.}
    \label{fig:dataflow}
\end{figure*}

The Bernoulli encoder uses a comparator to compare the input integer with a pseudo-random number (PRN) generated from an external source. Notably, we do not perform normalization in hardware. Instead, we directly compare the unnormalized input integer \( I \) with a random integer sampled uniformly from \( (0, I_{\text{max}}] \). In our design, \( I_{\text{max}} \) is set to \( d_K \) within SACs and to \( N \) outside of SACs. Choosing \( d_K \) and \( N \) as powers of two can streamline the hardware design by allowing directly using a fixed-point random integer sampled from a uniform distribution for the comparison. This approach simplifies the design and enables the use of power-efficient linear feedback shift register (LFSR)-based PRN generators.

\subsubsection{SSA Engine}
The SSA  engine is designed for efficient execution of the entire MHSA mechanism in spiking transformers. It consists of multiple SSA tiles, each serving as an individual attention head. These tiles are stateless, allowing them to be reused across different layers of the transformer, enhancing both area and energy efficiency. Additionally, the SSA engine incorporates an LFSR array that generates all the necessary PRNs for the Bernoulli neurons, ensuring a consistent supply of random numbers required for the stochastic elements of the model.

To further optimize power and area efficiency, the SSA engine implements a custom reuse strategy for random number generation, as described in \cite{katti2024bayesian}. Here we implement a 32-bit LFSR and tap all 4 bytes from the LFSR instead of just the lowest 8-bits \cite{AnalogDevices}. This approach maximizes the utilization of generated random numbers, conserving both power and area. By leveraging these design strategies, the SSA engine achieves a balance of high performance and resource efficiency.

\subsection{Dataflow}

Fig.~\ref{fig:dataflow} illustrates the inference data flow of Xpikeformer, featuring the alternating operation of the AIMC engine and the SSA engine, with data exchange facilitated through on-chip SRAM.

The AIMC engine serially drives input data through each forward propagation or fully connected layer. Each AIMC cycle, except for the first and last, includes two feedforward layers and one fully connected layer to project embeddings into the $QKV$ feature space. In each layer, data is processed in a \textit{token-wise event-driven} manner: the spike encoding of each token embedding is input sequentially over $T$ time steps, followed by cycling through $N$ tokens. This approach allows the membrane potential stored in the LIF unit's register to accumulate directly and reset upon token switching. Looping $N$ tokens for each time step $t$ would require storing and reading back membrane potentials, leading to unnecessary data transmission overhead.

Between layers within the same AIMC engine, $N$ token embeddings are forwarded in a pipelined manner. During the forward transmission of each token embedding, the spike encoding over $T$ time steps is also pipelined between layers.

The SSA engine operates multiple SSA tiles in parallel, with each tile handling one head of the MHSA mechanism. As previously discussed, queries, keys, and values in SSA tiles are transmitted in a \textit{matrix-wise event-driven} manner. Specifically, the three matrices $Q^t$, $K^t$, and $V^t$ are streamed to the SSA tile column by column over $d_K$ clock cycles. The subsequent matrices $Q^{t+1}$, $K^{t+1}$, and $V^{k+1}$ are transmitted in the next $d_K$ clock cycles. When pipelining in time steps, the computational delay of an SSA tile (from first input to first output) is approximately $d_K$ clock cycles.

\section{Training and Inference Details}\label{se_train}

\subsection{Integrated Framework for Hybrid Training}\label{se1_hwat}

To effectively train Xpikeformer, we adopt a hybrid framework that integrates \textit{SpikingJelly} \cite{fang2023spikingjelly} and the \textit{IBM Analog Hardware Acceleration Toolkit (AIHWKit)} \cite{rasch2021flexible}. SpikingJelly defines and manages the spiking components of Xpikeformer, particularly the LIF neurons and the SSA. AIHWKit is adopted to simulate the nonidealities inherent in AIMC.

We adopt a two-stage training procedure: 
\begin{enumerate}
    \item Conventional training (CT) is first performed in an ideal full-precision environment. 
    \item The trained parameters are then fine-tuned using hardware-aware training (HWAT), where noise is injected during forward propagation to simulate PCM-based hardware imperfections, while backward propagation remains ideal.
\end{enumerate}

During training, forward propagation outputs are averaged over time steps to compute the loss, which follows a definition similar to that used in ANN transformer training. The network is trained using the AdamW optimizer  \cite{kingma2014adam}.

\subsection{Global Drift Compensation}\label{se1_gdc}
To mitigate conductance drift in PCM devices within the AIMC engine, we implement a global drift compensation (GDC) method based on \cite{joshi2020accurate}. During the calibration phase, Xpikeformer periodically measures the output current of several SA columns using a known calibration input voltage. This allows Xpikeformer to track global shifts in conductance. During inference, the outputs are scaled by a factor derived from these calibration measurements. Calibration can be performed while tiles are idle, with current summation implemented within the system's control unit to minimize overhead.

\section{Accuracy Evaluation}\label{se_acc}
The inference accuracy of Xpikeformer is validated in hardware-simulated environments using AIHWKit. We evaluate the encoder-only and decoder-only architectures separately, each through a dedicated task.
\subsection{Accuracy Results}\label{se1_acc}

\emph{Task 1 (Image Classification): }
To evaluate the accuracy performance of the encoder-only transformer, we select the image classification task, as it requires the model to extract and process relevant features from the entire input image in parallel. We train three different-sized vision transformers (ViTs) using the Xpikeformer implementation \textit{from scratch} on the CIFAR10 and ImageNet-1K datasets, following the training strategy introduced in Section \ref{se1_hwat}. Inference accuracy is tested on the evaluation dataset using the simulation platform described in Section~\ref{se1_gdc}. We compare the Xpikeformer results against two baseline implementations: (\emph{i}) \textit{ANN-ViT (GPU)} -- Vanilla ViT trained with mixed precision (FP16/32) on GPU \cite{dosovitskiy2021image} and (\emph{ii})  \textit{SNN-ViT (GPU)} -- SOTA software design of SNN-based ViT trained with FP16/32 on GPU \cite{zhou2022spikformer}. The parameters of both baselines are INT8-quantized during the test stage.  

The accuracy results are summarized in Table~\ref{tab:imageacc}. For SNN-based models, the minimum spike encoding length $T$ required for convergence (defined as $\Delta\text{Acc}<0.1$) is indicated in brackets. Xpikeformer generally requires a longer spike encoding length to converge compared to the \textit{SNN-ViT (GPU)} due to hardware noise and stochasticity introduced by Bernoulli neurons. Despite this, Xpikeformer achieves competitive results, attaining 84.74\% on CIFAR-10 and 70.50\% on ImageNet, with accuracy gaps of less than 1.5\% compared to same-sized ANN-ViT models.
 
While Xpikeformer consistently requires a longer spike encoding length than \textit{SNN-ViT (GPU)} for the same task, the additional encoding overhead is smaller on ImageNet compared to CIFAR-10. This is because transformers generalize poorly on limited data but exhibit greater robustness to hardware noise and stochasticity when trained on larger, more diverse datasets. 

Scaling up the model size further improves Xpikeformer's performance while reducing the required spike encoding length. For instance, increasing the model size from 6-512 to 8-768 reduces the spike encoding length requirement from 8 to 7, while the accuracy gap relative to \textit{ANN-ViT} decreases from 2.22\% to 1.14\%. This improvement occurs because the larger model dimensions effectively average out hardware noise and Bernoulli coding uncertainties.

\begin{table}[t]
    \centering
    {
    \caption{Accuracy Performance on Image Classification Tasks}
    \begin{tabular}{cccc}
    \toprule 
    \multirow{2}{*}{Model}&\multirow{2}{*}{\makecell[cc]{Size /\\Depth-Dim}}  & CIFAR10   & ImageNet-1K \\ 
    &  & Accuracy ($T$)  & Accuracy ($T$) \\ 
    \toprule 
    \multirow{3}{*}{\makecell[cc]{ANN-ViT\cite{dosovitskiy2021image}\\(GPU)}}&4-384  &  83.41 & - \\
    &6-512 & 86.02 & 65.58 \\
    &8-768 & - & 71.64 \\
    \hline
    \multirow{3}{*}{\makecell[cc]{SNN-ViT\cite{zhou2022spikformer}\\(GPU)}}&4-384  &  82.83 (5) & -\\
    &6-512  & 85.68 (4) & 65.42 (6) \\
    &8-768 & - & 71.13 (4) \\
    \hline
    \multirow{3}{*}{\makecell[cc]{Xpikeformer-ViT\\(Simulated ASIC)}}&4-384  & 82.66 (11)  &-  \\
    &6-512  & 84.74 (10) & 63.36 (8) \\
    &8-768 & - & 70.50 (7) \\ \toprule 
    \end{tabular}
    \label{tab:imageacc}
    }
\end{table}

\begin{table}[t]
    \centering
    {
    \caption{Accuracy Performance on In-Context Learning for Wireless Symbol Detection}
    \begin{tabular}{cccc}
    \toprule 
    \multirow{2}{*}{Model}&\multirow{2}{*}{\makecell[cc]{Size /\\Depth-Dim}}  &  $2\times2$ Antennas &  $4\times4$ Antennas  \\ 
    &  & BER ($T$)  & BER ($T$) \\ 
    \toprule 
    \multirow{2}{*}{\makecell[cc]{ANN-GPT\cite{zecchin2023context}\\(GPU)}} &4-256  &  0.051 & 0.141 \\
    &8-512 & 0.049 & 0.077 \\
    \hline
    \multirow{2}{*}{\makecell[cc]{SNN-GPT\cite{song2024neuromorphic}\\(GPU)}}&4-256  & 0.061 (6)  & 0.196 (7) \\
    & 8-512  & 0.058 (4) & 0.086 (4) \\
    \hline
    \multirow{2}{*}{\makecell[cc]{Xpikeformer-GPT\\(Simulated ASIC)}}&4-256  & 0.067 (6)  & 0.205 (11)  \\
    & 8-512  & 0.063 (4) & 0.091(5) \\ \toprule 
    \end{tabular}\label{tab:iclacc}
    }    
\end{table}

\emph{Task 2 (In-Context Learning for Wireless Communication):}  
Decoder-only transformers are well-suited for tasks involving temporal sequential dependencies, particularly demonstrating the ability of in-context learning (ICL). In ICL, the model is presented with a sequence of query-answer pairs following a consistent pattern. Without explicit training on this pattern, it predicts the answer to a new query by attending to preceding query-answer pairs within the input sequence. This capability has been shown to enable wireless receivers to accurately classify symbols, even when the signals have passed through unknown wireless channels and been corrupted by nonlinearities \cite{zecchin2023context,rajagopalan2023transformers}.

Following the formulation in \cite{song2024neuromorphic}, we adopt the two-symbol detection task, assuming 2$\times$2 and 4$\times$4 antenna configurations. A larger number of antennas implies higher computational complexity. Performance is evaluated using the bit error rate (BER) metric, where a lower BER indicates higher detection accuracy. The number of input query-answer pairs is fixed at 18. We compare the performance of Xpikeformer with the following baselines: (\emph{i}) \textit{ANN-GPT (GPU)} -- a GPT-2-like decoder-only transformer running on GPUs \cite{zecchin2023context}, and (\emph{ii}) \textit{SNN-GPT (GPU)} -- the SOTA software design of an SNN-based solution running on GPUs \cite{song2024neuromorphic}. The baseline models are trained using FP16/32 precision and evaluated with INT8 quantization during the test phase.

The results are summarized in Table~\ref{tab:iclacc}. On the 2$\times$2 antenna task, both 4-256 and 8-512 Xpikeformers achieve a low BER using the same spike encoding length ($T=6$ and $T=4$, respectively) as \textit{SNN-GPT (GPU)}. This is because the task, characterized by a small number of classes, is relatively simple. However, as the number of antennas increases, the number of classes grows exponentially, leading to degraded performance for both small-sized \textit{SNN-GPT} and Xpikeformer-GPT models. Scaling up Xpikeformer-GPT from 4-256 to 8-512 brings greater performance gains than \textit{SNN-GPT}, reducing BER by 0.114 with 6 fewer time steps, compared to a 0.110 reduction with 3 fewer time steps for \textit{SNN-GPT}. This improvement is attributed to the larger model's ability to better learn and average out hardware noise and stochastic variations.

\subsection{Ablation Study: Long-term Accuracy Performance}
\begin{figure}[t]
    \centering
    \includegraphics[width=0.9\linewidth]{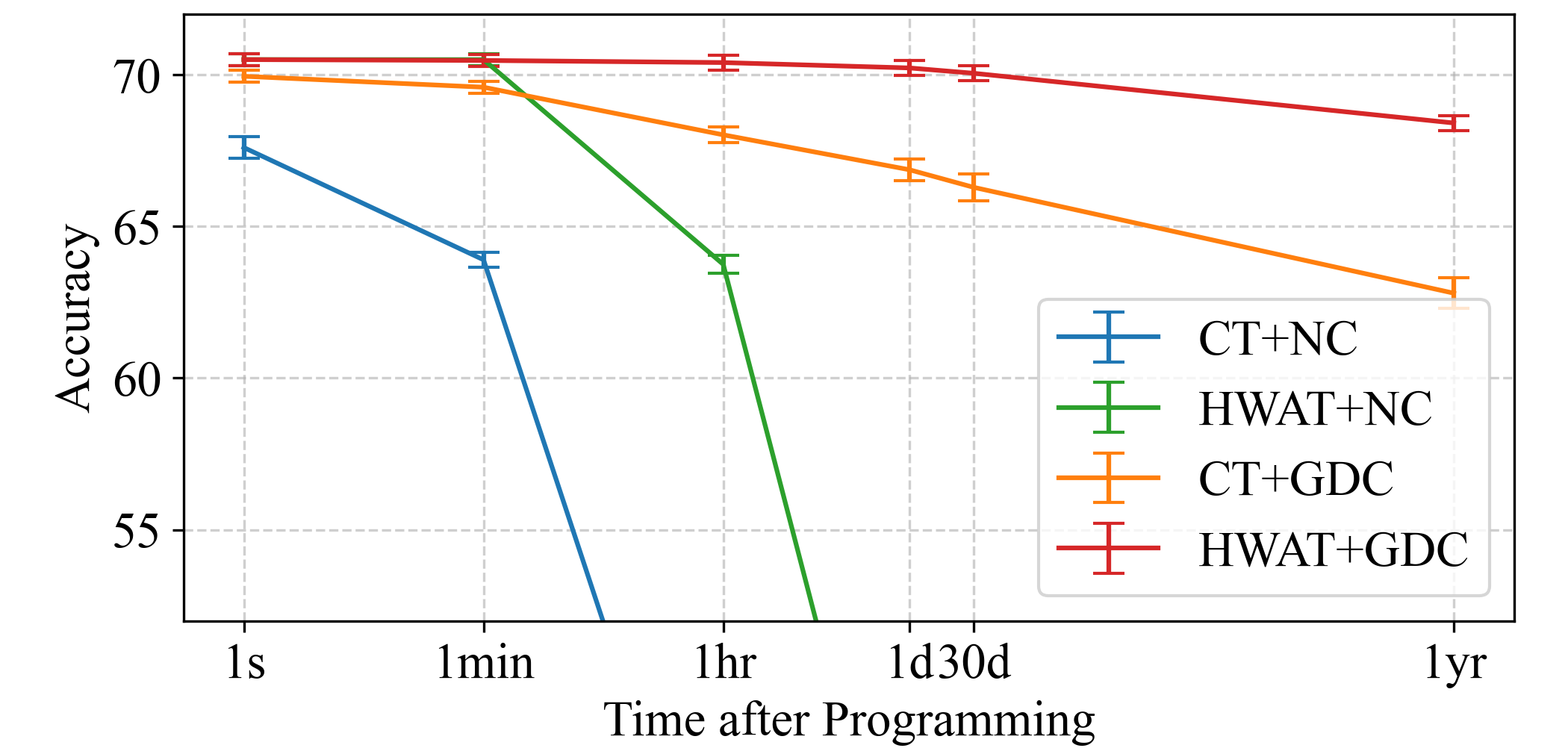}\label{fig_acc_drift}
    {
    \caption{Long-term ImageNet accuracy of Xpikeformers trained with different strategies and drift compensation methods: conventional training (CT) + no compensation (NC), CT + global drift compensation (GDC), hardware-aware training (HWAT) + NC, and HWAT+GDC. }
    }
\end{figure}
To evaluate the impact of NVM conductance drift on performance, we analyzed Xpikeformer's accuracy over time under different training strategies and drift compensation methods. Four Xpikeformer-ViTs of size 8-768 were trained using all combinations of two training strategies—CT and HWAT—and two calibration settings—with and without GDC.

Without GDC or HWAT, accuracy started at 67.6\% but dropped significantly to below 50\% within one hour. With HWAT but no GDC, accuracy remained above 70\% initially but rapidly declined to below 65\% within an hour due to conductance drift. With GDC alone, accuracy improved to 69.9\% and remained above 60\% for one year. Combining HWAT and GDC achieved the best stability, with accuracy dropping by only 2.09\% after one year. Similar trends were observed for other tasks tested in Section~\ref{se1_acc}.

We also compared the long-term accuracy of Xpikeformer-ViTs (6-512 vs. 8-768), as shown in Table~\ref{tab:longterm}. Despite having fewer parameters, the 6-512 model exhibits a larger accuracy drop ($-$2.95\% after one year) compared to the 8-768 model. The smaller model also gained less benefit from GDC alone. This is because larger model dimensions help average out hardware noise uncertainties and the stochastic component of conductance drift during GDC calibration.

\begin{table}[t]
    \centering
    {
    \caption{Xpikeformer Long-Term (One Year) Accuracy on ImageNet-1K}
    \begin{tabular}{ccccc}
    \toprule 
    Size & CT+NC &  HWAT+NC &  CT+GDC &  HWAT+GDC \\ 
    \toprule 
    6-512  & 9.63 (-56.73) & 11.32 (-55.04) & 54.37 (-11.99) & 60.41 (-2.95)\\
    \hline
    8-768  & 6.10 (-64.40) & 5.46 (-65.04) & 62.80 (-7.70) & 68.41 (-2.09)\\
    \toprule 
    \end{tabular}
    }
    \label{tab:longterm}
\end{table}

\section{Efficiency Evaluation}\label{se_energy}

\subsection{Energy Consumption} We evaluate the runtime energy consumption of the entire network in two parts: \textit{computational energy} and \textit{runtime memory access} energy consumption.
{
\subsubsection{Baselines} We compare the runtime energy consumption of Xpikeformer with three baseline architectures:
 \textbf{(\emph{i}) \textit{ANN-Quant}} -- a SOTA fully digital accelerator for ANN-based transformer models, as reported in \cite{marchisio2023swifttron}. This baseline serves as a benchmark for fully digital ANN architectures. \textbf{(\textit{ii}) \textit{ANN-Quant+AIMC}} -- this baseline modifies (\emph{i}) by replacing the feed-forward and fully connected layers with AIMC crossbars based on PCM, reducing energy costs associated with MAC operations. \textbf{(\emph{iii}) \textit{SNN-Digi-Opt} } -- this baseline assumes an ideal digital implementation of spiking transformer using the arithmetic operations proposed in \cite{zhou2024spikformer}. 
 }
\subsubsection{Evaluation Methods}
For the AIMC crossbars used in both Xpikeformer and the \textit{ANN-Quant+AIMC}, we customize the \textit{DNN+NeuroSim V1.4} framework \cite{peng2020dnn+} with the parameters specified in Table \ref{tab:parameters} to simulate the hardware implementations on a layer-by-layer basis. The area and energy estimates of the SSA tile were obtained by performing synthesis using the \textit{Cadence 45nm process design kit} (PDK).

To estimate the baseline energy consumptions, we consider all digital operations (MAC, AC, MUL, ADD, and logic gates) using energy metrics based on 45 nm CMOS technology as reported in \cite{pedram2017dark,buffa2021voltage}, following the approach in \cite{acesnn2022}. 
We assume that there is sufficient on-chip cache to keep the model weights loaded for the entire duration of the hardware operation. Therefore, the energy consumption for loading model parameters is excluded from our runtime memory access energy calculations. For all memory access operations required to read or write input data, output data, and intermediate results, we assume that all necessary data is read from or written to the on-chip SRAM. For all input data, output data, and intermediate results in ANNs, as well as pre-activations in SNNs, we assume they are quantized to INT8 and stored in 8-bit SRAM when necessary. For the fully digital configurations \textit{ANN-Quant} and \textit{SNN-Digi-Opt}, we assume all model parameters are also quantized to INT8 and pre-loaded into the cache. 

{ For SNN implementations, as both computing energy and runtime memory access energy are directly proportional to the spike encoding length \( T \), we use the minimum required spike encoding length \( T \) for Xpikeformer and \textit{SNN-Digi-Opt}, respectively, as determined in Section~\ref{se1_acc}, to achieve their converged accuracy. This ensures a fair comparison of computational efficiency, as both per-inference energy consumption and latency scale directly with the spike encoding length in Xpikeformer and the baseline hardware implementations considered.}

\subsubsection{Results}
\begin{figure}[t]
    \centering
    \subfigure[]{\includegraphics[width=0.48\linewidth]{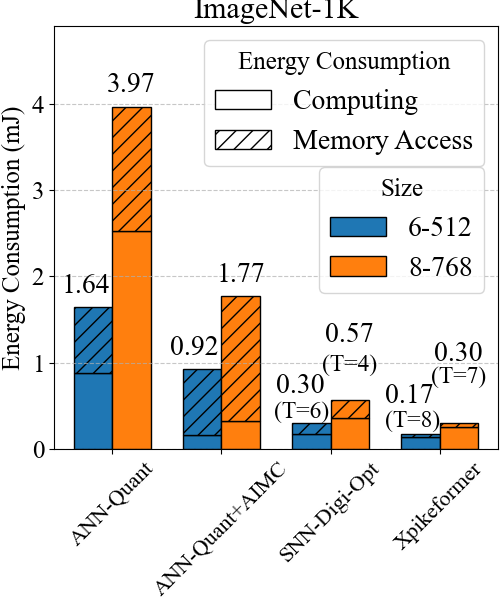}\label{fig_energy_task1}}
    \subfigure[]{\includegraphics[width=0.48\linewidth]{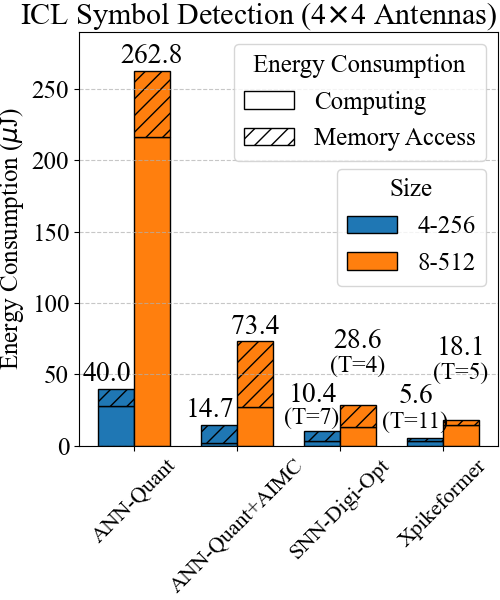}\label{fig_energy_task2}}
    {
    \caption{Energy consumption comparison between Xpikeformer and baseline implementations on downstream application (a) ImageNet-1K and (b) ICL symbol detection ($4\times4$ antennas).}
    \label{fig:energy_1}
    }
\end{figure}

We evaluate the energy consumption on both Task 1 (ImageNet) and Task 2 (4$\times$4 antennas) across different model sizes. The energy consumption comparisons between Xpikeformer and the baseline implementations are shown in Fig.~\ref{fig:energy_1}. It can be observed that Xpikeformer consistently consumes less energy than all three baseline models across varying model sizes. Specifically, on ImageNet-1K, Xpikeformer achieves 9.6-13$\times$ energy reduction than \textit{ANN-Quant} and 5.4-5.9$\times$ energy reduction than \textit{ANN-Quant+AIMC}. Notably, despite using more timesteps to achieve comparable accuracy, Xpikeformer still consumes 1.8-1.9$\times$ less energy than \textit{SNN-Digi-Opt}. Similar results go for the ICL symbol equalization task, where Xpikeformer achieves energy reduction 7-14.5$\times$ than \textit{ANN-Quant}, 2.6-4.0$\times$ than \textit{ANN-Quant+AIMC}, and 1.6-1.8$\times$ than \textit{SNN-Digi-Opt}. As the model scales up, the energy efficiency becomes more evident due to the reduced requirement for spike encoding length \( T \).

Several key insights can be drawn from comparing computational energy consumption. Firstly, the results for \textit{ANN-Quant} demonstrate that MAC operations are highly energy-hungry, usually dominating more than 90\% computing energy even when operating at low INT8 precision. This energy overhead represents a bottleneck that traditional quantized ANN-based transformer accelerators cannot overcome. In contrast, the other implementations address this challenge using different approaches: \textit{ANN-Quant+AIMC} and Xpikeformer leverage AIMC to achieve efficient MVM in fully connected and linear layers, while \textit{SNN-Digi-Opt} replaces MAC operations with masked ADD operations, taking advantage of the binary nature of SNN activations. Additionally, both SNN implementations avoid energy-intensive operations such as softmax and layer normalization, which remain necessary in ANN accelerators.

Although Xpikeformer requires a few additional timesteps, it still consumes less computational energy than \textit{SNN-Digi-Opt}. This advantage stems from two key factors:  
(1) For the attention mechanism, \textit{SNN-Digi-Opt} uses masked addition and integer multiplication to compute attention scores, whereas Xpikeformer employs a more energy-efficient logic-gate-based SSA engine.  
(2) For feed-forward layers, while AIMC requires complex peripheral circuitry and ADCs, but is generally more energy-efficient per timestep than the digital masked addition approach. Specifically, AIMC computes MVM with $O(1)$ complexity, whereas the digital method requires $O(n)$ operations, with $n$ being the dimension of the vector.

\begin{figure}[t]
    \centering
    \includegraphics[width=0.9\linewidth]{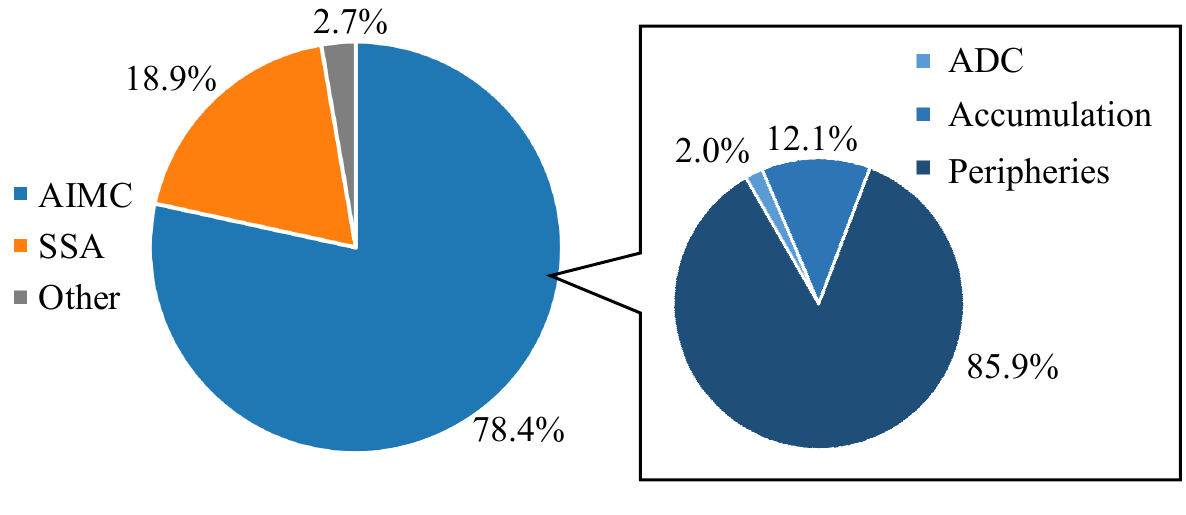}
    \caption{Breakdown of Xpikeformer computational energy.}
    \label{fig:energy_breakdown}
\end{figure}

From the runtime memory access perspective, we observe that both \textit{ANN-Quant} and \textit{ANN-Quant+AIMC} consume the same high amount of memory access energy, as AIMC does not reduce intermediate data storage overhead. In contrast, both SNN implementations achieve lower memory access energy compared to ANN accelerators, primarily owing to the softmax-free SNN attention mechanism and the binary nature of SNN activations. However, this memory efficiency benefit diminishes when the number of timesteps \( T \) exceeds the bit width of the ANN activation datatype.

Xpikeformer achieves significantly lower memory access energy compared to \textit{SNN-Digi-Opt}, because 
(1) \textit{SNN-Digi-Opt} requires storing \( T \) non-binary pre-activations from every crossbar output column before performing LIF calculations, whereas Xpikeformer’s rowblock-wise mapping strategy in its AIMC engine eliminates this bottleneck.  
(2) The streamlined design of the SSA engine eliminates the need to store intermediate attention scores.

Fig.~\ref{fig:energy_breakdown} presents the breakdown of Xpikeformer computational energy consumption. The majority of the computing energy is consumed by the AIMC engine, accounting for 78.4\% of the total computational energy. The SSA engine contributes 18.9\%, while other components, such as residual units, collectively account for 2.7\%. A more detailed breakdown of the AIMC engine's energy consumption is shown on the right side of Fig.~\ref{fig:energy_breakdown}. Notably, the ADC consumes only 2.0\% of the AIMC energy. This efficiency is achieved by sharing ADCs across multiple read-out channels through multiplexing (Xpikeformer uses a sharing ratio of 8, balancing energy efficiency and latency), which significantly reduces static power consumption that would otherwise be incurred by using dedicated ADCs for each channel. We also observe that periphery circuits (decoders, multiplexers, switch matrices, buffers, etc.) consume 85.9\% of the AIMC energy. This is because ADC multiplexing requires a dedicated decoder and control logic, making these components relatively more energy-intensive than crossbar itself. Additionally, the SA mapping strategy introduces additional control overhead between shared SRAM and local SA buffers. This overhead is unavoidable in this work due to the crossbar size limitations discussed in Section \ref{se1_aimcengine}. Finally, accumulation processes, including differential adders and LIF units, which involve digital operations, contribute 12.1\% of AIMC energy.

\subsection{Latency and Area Evaluation}
\begin{figure}[t]
    \centering
    \subfigure[]{\includegraphics[height=4.5cm]{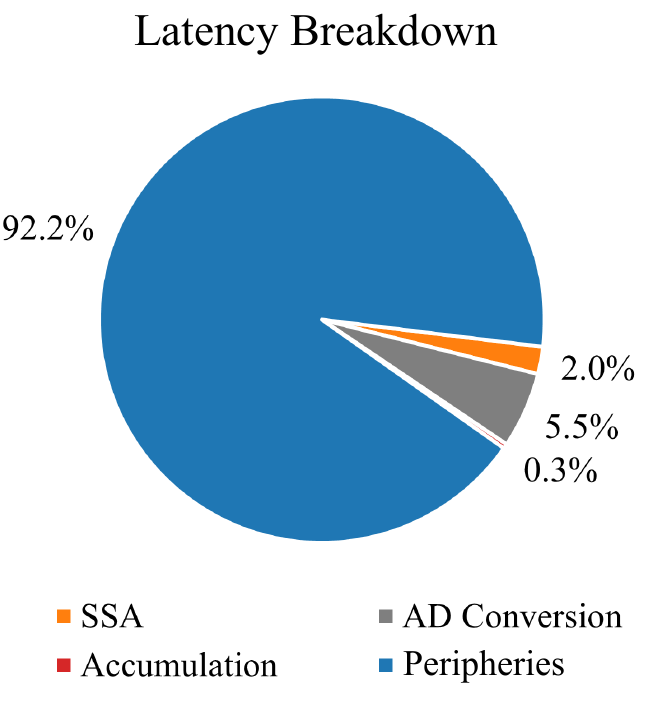}\label{fig:latencybreakdown}}\hspace{1cm}
    \subfigure[]{\includegraphics[height=4.5cm]{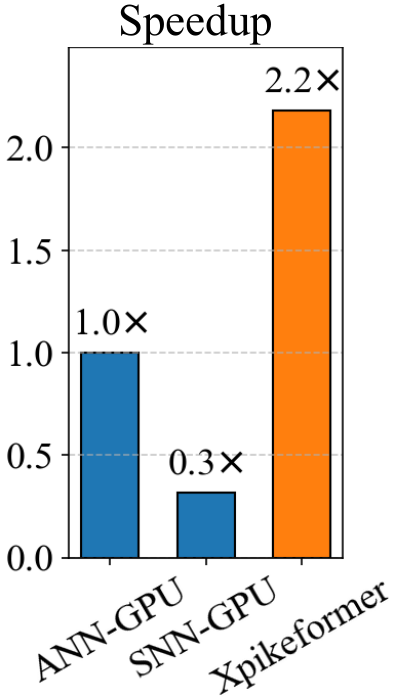}\label{fig:latencyefficiency}}
    \caption{(a) Latency breakdown of Xpikeformer and (b) Latency comparison with ANN and SNN transformers on GPU.}
\end{figure}

The latency breakdown of Xpikeformer is shown in Fig.~\ref{fig:latencybreakdown}. More than 92\% of the latency is attributed to peripheral circuit operations, as these handle global data movement, routing, and control tasks, which are inherently sequential and cannot be parallelized like core computations. Additionally, decoding, buffering, and interconnect operations scale with model size, further contributing to the overhead. In contrast, the actual AIMC computing latency is negligible, accounting for only 0.3\% of the total latency. The SSA computation contributes only 2.0\% of the overall latency.

We compare the per-inference latency of Xpikeformer with two same-sized baseline implementations: an ANN-based transformer and spiking transformer as in \cite{zhou2024spikformer}, both deployed on a GPU (Nvidia RTX A2000), adopting the minimum spiking encoding lengths as in Section~\ref{se1_acc}. Results are shown in Fig.~\ref{fig:latencyefficiency}. Despite the extra temporal dimension, Xpikeformer achieves a $2.18\times$ speedup over the GPU implementation of ANN-based transformers, benefiting from fast in-memory MVM and the parallel, streamlined SSA design. Meanwhile, the spiking transformer suffers from poor GPU efficiency due to precision mismatches and lack of hardware optimization. In contrast, Xpikeformer delivers a 6.85$\times$ speedup over it.

The area of Xpikeformer's computing engines is estimated using the \textit{DNN+NeuroSim V1.4} framework and the Cadence synthesis tool at the 45 nm technology node. The total chip area of Xpikeformer is {784 mm\textsuperscript{2}}, with the majority (76.5\%) occupied by periphery circuits and interconnections. The AIMC computing core, which includes crossbars, ADCs, and accumulation units, accounts for {11.5\%} of the area. The SSA engine, being more area-efficient than traditional ALUs due to its use of {lightweight} logic gates as computing units, contributes {12\%} of the total area.

\subsection{Comparison With SOTA Hardware Accelerators}
To the best of the authors' knowledge, Xpikeformer is the first hardware accelerator designed for spiking transformers. We compare Xpikeformer with two SOTA ASIC accelerators for ANN-based transformers, summarized in Table~\ref{tab:comparison_state_of_the_art}. 

SwiftTron~\cite{marchisio2023swifttron} is a fully digital ASIC accelerator designed for quantized transformers. It replaces traditional floating-point ALUs with fixed-point computing units and approximates nonlinear operations using second-order polynomials. The reported chip area of SwiftTron is 2.9$\times$ smaller than that of Xpikeformer. However, this reduction is achieved by reusing computational resources through time multiplexing, which introduces additional energy consumption and latency due to frequent parameter reads during inference. Benchmarked on the same task, Xpikeformer achieves similar latency while consuming 13$\times$ less energy by leveraging AIMC for feed-forward layers and lightweight logic-gate-based computations for the attention mechanism.

X-Former~\cite{sridharan2023x} is an in-memory accelerator for ANN transformers that employs AIMC crossbars for feed-forward layers and DIMC for attention. Note that X-Former architecture uses 1-bit ReRAM for weight storage, which translates to lower weight storage density compared to Xpikeformer’s multi-bit PCM devices, potentially increasing the overall area for AIMC implementation. For MHSA computation, X-Former employs DIMC units based on SRAM arrays, which require writing $Q$ and $V$ matrices into SRAM crossbars during inference, and storing intermediate results in the memory, leading to additional latency. In contrast, Xpikeformer leverages the SSA engine to eliminate online memory-write operations and intermediate storage while also avoiding the complex peripherals associated with DIMC. Overall, Xpikeformer achieves a 6$\times$ reduction in inference energy and a 1.9$\times$ speedup over X-Former.

\begin{table}[t]
    \centering
    \fontsize{7pt}{8pt}\selectfont
    
    {
    \caption{Comparison With SOTA Accelerators}
    \label{tab:comparison_state_of_the_art}
    \begin{threeparttable}
    \begin{tabular}{lccc}
        \toprule
        \textbf{Performance Metric} & \textbf{\cite{marchisio2023swifttron}} & \textbf{\cite{sridharan2023x}} & \textbf{Xpikeformer} \\
        \midrule
        Paradigm           & ANN     & ANN         & SNN \\
        MAC Implementation & Digital ALU & ReRAM-AIMC & PCM-AIMC \\
        MHSA Implementation & Digital ALU & DIMC & SSA  \\
        Technology         & 65 nm   & 32 nm       & 45 nm \\
        Weight   & INT8     & INT8 (Equiv.)    & INT5 (Equiv.) \\
        Activation Precision  & INT8/32     & INT8     & Multi-Step Binary \\
        Frequency          & 143 MHz   & 200 MHz  & 200 MHz \\
        Area (mm$^2$)  & 273.0   & --   & 784  \\
        Energy/Inference* (mJ) & 3.97  & 2.04  & 0.30 \\
        Latency/Inference* (ms) & 2.26 & 4.13$^{\dagger}$ & 2.18$^{\dagger}$ \\
        \bottomrule
    \end{tabular}
        \begin{tablenotes}
            \fontsize{7pt}{8pt}\selectfont
            \item * Energy and latency results have been normalized to a consistent benchmark task (ImageNet-1K classification on ViT-8-768 with a patch size of \(16 \times 16\)) to ensure fair comparison across different architectures. For \cite{marchisio2023swifttron}, we assume their original chip area is fixed and scale the overall latency proportionally to the increase in task size. For \cite{sridharan2023x}, we assume that AIMC resources are sufficient to accommodate all model parameters, while the DIMC resources remain fixed as originally reported, with attention latency scaled up accordingly.
            \item $^{\dagger}$ The latency results for \cite{sridharan2023x} and Xpikeformer account for both computational latency and data movement from memory to computing cores.
        \end{tablenotes}
    \end{threeparttable}
    }
\end{table}

{
\subsection{Scalability of Xpikeformer}
For the SSA engine, separating SSA tiles from the LFSR array improves area efficiency by centralizing PRN generation in the shared LFSR array, leaving only basic logic gates and bit-wise counters in the tiles. In contrast, traditional fixed-point ALU arrays with adders and multipliers incur higher area overhead. The SSA engine's 1-bit I/O buses, with no intermediate storage or operations, ensure high data-dimensional flexibility. Its stateless, modular design supports both head-wise and layer-wise reuse, minimizing resource duplication and enabling efficient scaling to larger attention heads and additional layers without significant hardware overhead.

Like most NVM-based architectures, the AIMC engine in Xpikeformer scales spatially to accommodate larger weight matrices. By leveraging the multi-level programming capability of PCM devices, it achieves a 3$\times$ reduction in required memory devices compared to 1-bit ReRAM cells \cite{sridharan2023x}. With similar planar cell areas ($4F^2$–$8F^2$) to ReRAM, PCM crossbars provide denser weight storage, improving efficiency in device count and layout area for large-scale models. The row-block-wise mapping strategy further enhances memory utilization and weight distribution. For very-large models, Xpikeformer can scale further by connecting multiple nodes via chip-to-chip interconnects.
}

    
    

\section{Conclusion}\label{se_conclusion}


In this paper, we introduced Xpikeformer, a hybrid analog-digital hardware architecture designed to accelerate spiking neural network (SNN)-based transformer models. Xpikeformer integrates analog in-memory computing (AIMC) for feedforward and fully connected layers and a stochastic spiking attention (SSA) engine for efficient attention computation, mitigating the inefficiencies of deploying spiking transformers on general-purpose platforms.

Trained from scratch, spiking transformer models running on Xpikeformer achieve competitive accuracy on ImageNet-1K classification and wireless communication symbol detection tasks while maintaining high accuracy for over a year. Although the non-reusable nature of AIMC introduces an area tradeoff compared to non-AIMC digital accelerators implementing time multiplexing, Xpikeformer achieves 13$\times$ reduction in energy consumption at approximately the same throughput as the SOTA digital accelerator for ANN-based transformers.

In this work, we adopted NeuroSim's crossbar assumptions to facilitate energy, latency, and area evaluations. However, this approach may not fully capture the performance of SOTA AIMC technology. Practical implementations of NVM crossbars often adopt different input strategies, such as using input spikes to drive device conductivity through word lines and employing inverted voltage spikes and differential-input ADCs to represent positive and negative values in the crossbar \cite{8776540}. Future work will explore more energy-efficient NVM crossbar designs, advanced materials like 2D semiconductors, and real-world testing to further enhance Xpikeformer’s efficiency and applicability in energy-constrained environments.

\bibliographystyle{IEEEtran} 
\bibliography{references}

\end{document}